\documentclass[prb,aps,preprint,superscriptaddress,showpacs,floatfix]{revtex4}

\usepackage{graphicx}

\def\be{\begin{equation}}
\def\ee{\end{equation}}
\def\ba{\begin{eqnarray}}
\def\ea{\end{eqnarray}}

\def\124{YBa$_2$Cu$_4$O$_8$ }

\def\C60{A$_x$C$_{60}$ }

\begin{document}

\title{Scanning tunneling spectroscopy of SmFeAsO$_{0.85}$: Possible evidence for d-wave order parameter symmetry}

\author{Oded Millo}
\email{milode@vms.huji.ac.il} \affiliation{Racah Institute of
Physics, The Hebrew University of Jerusalem, Jerusalem 91904,
Israel}

\author{Itay Asulin}
\affiliation{Racah Institute of Physics, The Hebrew University of
Jerusalem, Jerusalem 91904, Israel}

\author{Ofer Yuli}
\affiliation{Racah Institute of Physics, The Hebrew University of
Jerusalem, Jerusalem 91904, Israel}

\author{Israel Felner}
\affiliation{Racah Institute of Physics, The Hebrew University of
Jerusalem, Jerusalem 91904, Israel}

\author{Zhi-An Ren}
\affiliation{National Laboratory for Superconductivity, Institute of Physics and Beijing National Laboratory for Condensed Matter Physics, Chinese Academy of Sciences, P. O. Box 603, Beijing 100190, P. R. China}

\author{Xiao-Li Shen}
\affiliation{National Laboratory for Superconductivity, Institute of Physics and Beijing National Laboratory for Condensed Matter Physics, Chinese Academy of Sciences, P. O. Box 603, Beijing 100190, P. R. China}

\author{Guang-Can Che}
\affiliation{National Laboratory for Superconductivity, Institute of Physics and Beijing National Laboratory for Condensed Matter Physics, Chinese Academy of Sciences, P. O. Box 603, Beijing 100190, P. R. China}

\author{Zhong-Xian Zhao}
\affiliation{National Laboratory for Superconductivity, Institute of Physics and Beijing National Laboratory for Condensed Matter Physics, Chinese Academy of Sciences, P. O. Box 603, Beijing 100190, P. R. China}

\begin{abstract}

We report a scanning tunneling spectroscopy investigation of polycrystalline SmFeAsO$_{0.85}$ having a superconducting transition at 52 K. On large regions of the sample surface the tunneling spectra exhibited V-shaped gap structures with no coherence peaks, indicating degraded surface properties.  In some regions, however, the coherence peaks were clearly observed, and the V-shaped gaps could be fit to the theory of tunneling into a \emph{d}-wave superconductor, yielding gap values between 8 to 8.5 meV, corresponding to the ratio $2\Delta$/$K_BT_C$ $\sim$ 3.55 - 3.8, which is slightly above the BCS weak-coupling prediction. In other regions the spectra exhibited zero-bias conductance peaks, consistent with a \emph{d}-wave order parameter symmetry.   
\end{abstract}


\maketitle

Among the main fundamental questions that naturally arose after the discovery of high temperature superconductivity (HTSC) in the oxy-pnictides family of layered Fe-based RFeAs(O$_{1-x}$,F$_x$) (R=rare-earth) materials,\cite{a1,a2} were the symmetry of the order parameter, the concomitant gap structure and the value of the energy gap. In spite of the considerable experimental and theoretical effort directed to resolve these issues, they are not yet fully established. While there are reports that present experimental evidence for a double-gap structure,\cite{a3} consistent with some theoretical predictions,\cite{a4} others show a single gap structure. \cite{a5} The resemblance between the oxy-pnictides system and the HTSC cuprates (e.g., the layered structure, the superconducting dome structure vs. doping, existence of spin-fluctuations) makes the \emph{d}-wave order-parameter scenario rather appealing. Indeed, various theoretical predictions for \emph{d}-wave superconductivity were published,\cite{a6,a7} while others support an unconventional \emph{s}-wave pairing.\cite{a8,a9} On the experimental side, various spectroscopic techniques, such as point-contact spectroscopy, high-resolution and angular-resolved photoemission spectroscopy, infrared reflectance spectroscopy and nuclear magnetic resonance, were applied in studies of different oxy-pnictides materials, yielding diverse results regarding both the symmetry of the order parameter and the value (or values) of the energy gap(s).  A conventional isotropic \emph{s}-wave order parameter was reported in Ref. 5, but more commonly the measurements suggested nodal \emph{p}-wave or \emph{d}-wave symmetries,\cite{a10,a11,a12,a13} in some cases also along with a double-gap structure.\cite{a3} The $2\Delta$/$K_BT_C$ ratio, indicative of the coupling strength, also varied considerably between the different reports, from close to the BCS weak-coupling value of 3.52,\cite{a10,a14} through an 'intermediate-coupling' regime of $\sim$4,\cite{a11} up to a value as high as $\sim$8.\cite{a15} It is thus obvious that that further experimental effort is needed in order to clarify the picture.  

Tunneling and Andreev-reflection spectroscopy are very suitable techniques for measuring the superconductor gap and determining the symmetry of the order parameter.\cite{a16} Until now, only point-contact spectroscopy results acquired mainly in the Andreev-reflection regime (namely, with highly-transparent junctions having low Z $(< 1)$ values in the Blonder-Klapwijk-Tinkham (BTK) model\cite{a17} and its extension to \emph{d}-wave superconductors by Tanaka and Kashiwaya\cite{a18}) were reported for the new family of Fe-As based superconductors. In some cases larger tunneling resistances (Z $\sim$ 2) were also used, but there is not yet any report of \emph{bone-fide} tunneling spectra, acquired with Z $\sim$ 5 tunnel junctions.  It is important to note that tunneling spectroscopy may provide different information on the superconductor properties than Andreev-reflection spectroscopy.\cite{a16,a19} The tunneling spectra monitor the quasi-particle excitation energy-gap and thus yield information on the pairing scale. The energy scale determined by Andreev reflection is associated, on the other hand, with the coherence energy range of the superconducting macroscopic quantum-condensate state.  While in conventional BCS superconductors these two energy scales coincide, there is ample evidence that the former energy scale exceeds the latter in underdoped high temperature superconductor cuprates.\cite{a19,a20} Due to the aforementioned possible resemblance between the cuprates and the oxy-pnictides, tunneling spectroscopy measurements are essential and could provide important complementary information to Andreev spectroscopy.  A very recent scanning tunneling spectroscopy (STS) investigation of in-situ cleaved (Sr$_{1-x}$K$_x$)Fe$_2$As$_2$ single crystals, a compound belonging to the related newly discovered Sr-122 family of superconductors, inferred unconventional order-parameter symmetry.\cite{a21} In this paper we present STS data on poly-crystalline (oxygen deficient) SmFeAsO$_{0.85}$. Our tunneling spectra suggest, in spite of the degraded surface quality whose effect cannot be ignored, that SmFeAsO$_{0.85}$ is a \emph{d}-wave superconductor in the weak-coupling limit. 

The poly-crystalline SmFeAsO$_{0.85}$ samples, in which doping is achieved via oxygen deficiency rather than by fluorine doping, were prepared by high pressure synthesis.\cite{a2} SmAs (pre-sintered), Fe and Fe$_2$O$_3$ were mixed together according to the nominal composition, then ground and pressed into pellets that were sealed in a BN crucible and sintered under 6 GPa at 1250 $^\circ$C for two hours. The purity and the tetragonal (S.G. P4/nmm) structure (a=3.897(6) \AA, c=8.407(1) \AA) were verified by powder X-ray diffraction (XRD) using MXP18A-HF type diffractometer. Previous $^{57}$Fe M\"{o}ssbauer measurements showed an existence of magnetic order along with superconductivity, which can be attributed to a foreign Fe-As (e.g., Fe$_2$As) phase that could not be disclosed by the XRD data.\cite{a22} The sample surface was polished just before mounting inside our home-made STM and cooling down to 4.2 K, at which all data were acquired.  Two samples were measured, yielding similar results. The surface morphology of the sample, featuring elongated crystallites, is depicted by the STM image in Fig. 1(a). The temperature-dependent magnetization curves in Fig. 1(b) (measured in a quantum Design SQUID magnetometer) manifest a rather sharp superconductor transition onset at 52 K. 

\begin{figure}
\includegraphics[width=3in]{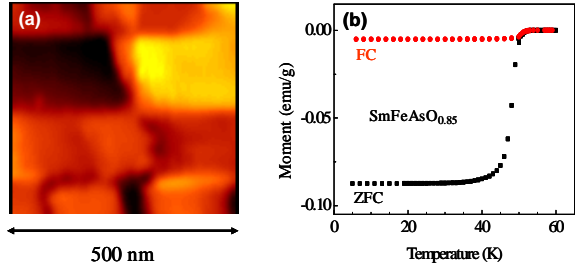}
\caption{(a) STM topographic image of a polycrystalline SmFeAsO$_{0.85}$ sample showing crystallites. (b) Zero field cooled (ZFC) and Field cooled (FC) temperature-dependent magnetization curves measured on the sample at 5 Oe.}\label{fig1}
\end{figure}

We now turn to discuss the typical energy-gap features, presented in Fig. 2. On most of the sample surface the tunneling spectra exhibited V-shaped gaps with no coherence peaks but with clear gap-edge features at $\pm$7 meV, as shown in Fig. 2(b). These structure resemble proximity-gaps observed on the normal side of HTSC/normal-metal junctions\cite{a23,a24} suggesting that superconductivity is suppressed on the sample surface and that the corresponding bulk superconductor gap is larger than 7 meV. In some regions, however, V-shaped gaps with clear coherence peaks were found, similar to tunneling spectra obtained on the (001) surface (\emph{c}-axis) of the HTSC cuprates [Fig. 2(a)]. The energy-gap values, $\Delta$, as determined from fits to the Tanaka and Kashiwaya model\cite{a18} for tunneling into a \emph{d}-wave superconductor (using a relatively small Dynes broadening parameter,\cite{a25} $\Gamma \sim 0.1\Delta$) ranged between 8 to 8.5 meV [red curve in Fig. 2(a)].  The corresponding $2\Delta$/$K_BT_C$ ratios are in the range of 3.55 - 3.8, close to that reported in the majority, but not all (e.g., Ref. 15) spectroscopic works on the oxy-pnictides.  We note that these values are slightly above the BCS weak-coupling value and smaller than those typically found for the cuprate HTSCs.  

\begin{figure}
\includegraphics[width=3in]{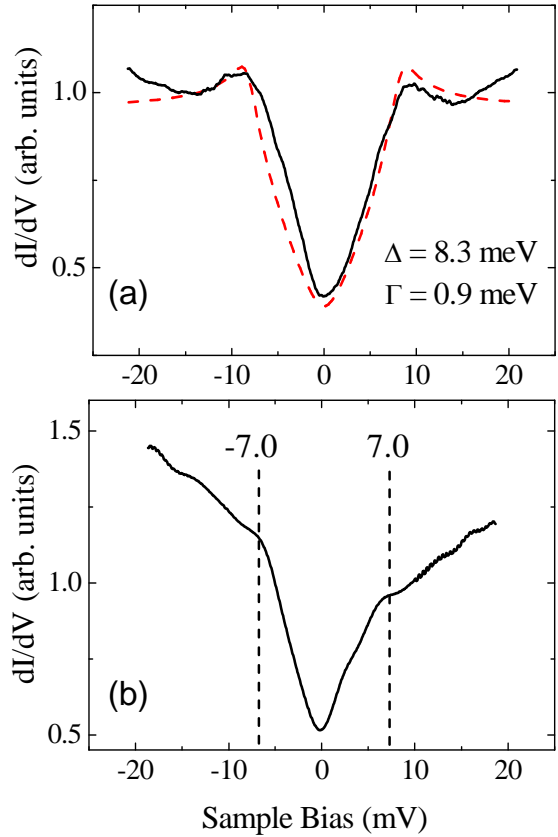}
\caption{Tunneling spectra at 4.2 K taken (a) in a region where coherence peaks were observed and (b) in a region where only sharp gap-edges were found.  The red dashed curve in (a) is a fit to the theory of tunneling into a \emph{d}-wave superconductor along the \emph{c}-axis, using the parameters denoted in the figure.}\label{fig2}
\end{figure}

\begin{figure}
\includegraphics[width=3in]{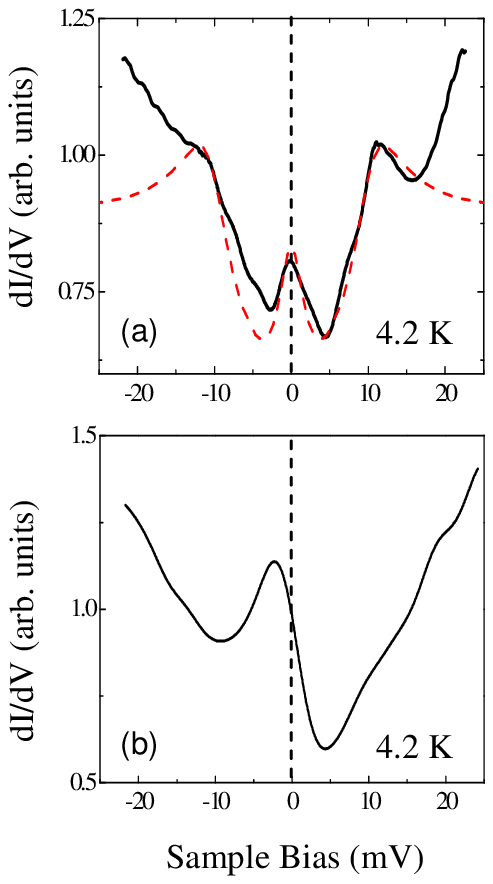}
\caption{(a) A tunneling spectrum exhibiting a zero-bias conductance peak with gap-like features. (b) A tunneling spectrum showing a shifted zero bias conductance peak. The red dashed curves are theoretical fits calculated as described in the text, and the vertical dashed lines are guides to the eye.}\label{fig3}
\end{figure}

The observed V-shaped gaps imply that SmFeAsO$_{0.85}$ has a nodal order parameter symmetry, either \emph{p}-wave or \emph{d}-wave. It is well established that the latter gives rise to Andreev bound states on the nodal surface of the superconductor, which manifest themselves by the emergence of a zero bias conductance peak (ZBCP) in tunneling spectra taken in the nodal direction, and (in a less pronounced way) also in spectra acquired in other non anti-nodal directions.\cite{a18} We have observed such ZBCPs within gap-like features on some locations, as exhibited by the spectrum presented in Fig. 3(a). This spectrum resembles STS spectra measured on the cuprates (e.g., see Ref. 26), and its main features can be reasonably fit using the \emph{d}-wave superconductor tunneling model, assuming in-plane tunneling at an angle of 15$^\circ$ with respect to the nodal direction. However, unlike our previous invetigation \cite{a26} of YBa$_2$Cu$_3$O$_{7-\delta}$, we could not find a correlation between the local surface morphology and the tunneling spectra.  Moreover, the abundance of the ZBCPs was smaller compared to the case of the hole-doped cuprates HTSCs.  This may be due to surface disorder that obstruct the multi-Andreev reflection process needed for the formation of zero-energy surface Andreev bound states,\cite{a27,a28} an issue that was raised also by Yates et al. in the discussion of their point-contact spectra.\cite{a29} Another interesting feature we have occasionally observed was a shift of the peak to a finite (small) voltage, as shown in Fig. 3(b). Such shifts may be related to the (probably foreign-phase induced) local magnetic order described above,\cite{a22} reminiscence of our previous finding on SrRuO$_3$-YBa$_2$Cu$_3$O$_{7-\delta}$ ferromagnetic-superconducting bilayers.\cite{a30}

In summary, our scanning tunneling spectroscopy data suggest that SmFeAsO$_{0.85}$ is a weak-coupling \emph{d}-wave superconductor. However, measurements on freshly cleaved single crystal samples are still needed in order to further verify this conclusion.

Acknowledgement: This work was supported by the Israel Science Foundation.

\end{document}